\documentstyle[prc,aps,epsf]{revtex}
\tightenlines
\begin{document}
\draft
\preprint{MKPH-T-97-19}
\title{%
Relativistic Effects and
the Role of Heavy Meson Exchange in Deuteron Photodisintegration%
\thanks{Supported by the Deutsche Forschungsgemeinschaft (SFB 201).}%
\footnote{Dedicated to Prof.\ J.A.\ Tjon on the occasion 
of his 60th birthday.}}
\author
{F.\ Ritz and
 H.\ Arenh\"ovel}
\address{%
 Institut f\"ur Kernphysik, 
 Johannes Gutenberg-Universit\"at, 
 D-55099 Mainz, Germany.}
\author
{T.\ Wilbois}
\address{%
 Institut f\"ur Theoretische Physik, 
 Universit\"at Hannover, 
 D-30167 Hannover, Germany.}
\maketitle
\begin{abstract}
  Relativistic effects and
  the role of heavy meson exchange in deuteron
  photodisintegration are studied systematically for
  photon energies
  below the 
  pion production threshold.
  In a $(p/M)$-expansion, all leading order
  relativistic one-body and $\pi$-exchange
  as well as all
  static heavy meson exchange currents consistent with the Bonn
  OBEPQ model are included. 
  In addition, one- and two-body boost effects
  have been investigated.
  Sizeable effects from the various two-body contributions
  beyond $\pi$-exchange have been found 
  in almost every observable considered, i.e.,
  differential cross section and single polarization observables.
\end{abstract}


\section{Introduction}
\label{introduction}

Electromagnetic deuteron disintegration is one of the basic processes in order
to study various aspects of the strong interaction of nucleons in nuclei.
For example, a large number of experimental and theoretical papers have
clarified the role of pion degrees of freedom as manifest in meson exchange
currents (MEC)
and the importance of relativistic contributions \cite{ArS91}.
However, the role of heavy meson exchange (with exception of the $\rho$ meson),
which give an important contribution to the $NN$ interaction remained largely
unclear. One of the reasons lies in the Siegert theorem which provides via
the Siegert operators in conjunction with the Siegert hypothesis a convenient
calculational tool to include implicitly the major part of MEC in the
electric multipoles \cite{Sie37}.

But admittedly such a procedure overshadows
the underlying physics, and only ``patches'' an inconsistent calculation.
A calculation that uses a set of consistent electromagnetic
(e.m.) operators with respect to
the hadronic interaction model should always be preferred.
Such a consistent treatment has been given some time ago allowing only pions
to interact with nucleons \cite{GoA92}. In particular, the importance of
consistency of the leading order relativistic contributions has been pointed
out in this work. However, the extension to a realistic interaction model was
missing.
Only recently, Levchuk has presented a nonrelativistic calculation of
deuteron photodisintegration for the Bonn OBEPR model
where all heavy meson exchange currents were included explicitly
so that Siegert operators were not needed \cite{Levchuk}.
Unfortunately the results are presented in such a way, that the specific
contributions from heavy meson exchange are not evident.
Furthermore relativistic contributions have been neglected,
which however become increasingly important at higher energies.
As a sideremark, we would like to mention another earlier calculation based
on the Bonn potential models with a consistent pion exchange current but
contributions of heavier mesons were included via the Siegert operators
only \cite{kms,blanpied}. 

More recently, we have extended the work of \cite{GoA92} in order to
investigate this question for deuteron 
electrodisintegration \cite{FR97}
taking as interaction model the Bonn OBEPQ versions \cite{MHE87,Mac89}.
As general result we found that the $\rho$ meson gives the most important 
heavy meson contribution whereas the influence of $\eta$, $\omega$, $\sigma$,
$\delta$, and $\gamma\pi\rho/\omega$ is much smaller, in some observables
completely negligible, in particular, near the quasifree kinematics.
However, it is a priori not clear that this conclusion will be valid also for
photodisintegration because of the fixed energy-momentum transfer relation.
Therefore, we want to provide with the present work the missing study of the
influence of heavy meson exchange on deuteron photodisintegration with
inclusion of competing relativistic effects in the one-body and pion exchange
sector.

The calculational framework is the same as in \cite{FR97} and will be very
briefly reviewed in Sect.~\ref{framework}\@. 
The results are presented and discussed in Section~\ref{results} 
restricting ourselves to the 
differential cross section $d\sigma/d\Omega$, 
and all single polarization observables, i.e.,
the photon asymmetry $\Sigma^l$, 
the target asymmetries $T_{IM},\; IM\in\{11,20,21,22\}$,
and the proton and neutron polarization components $P_y(p)$ and $P_y(n)$.
Section~\ref{summary} gives a short summary and outlook.

\section{Theoretical Framework}
\label{framework}

The calculation of the photodisintegration process is based on the
equations-of-motion method for the derivation of the hadronic interaction
model and the corresponding electromagnetic current operators.
It is described in detail in \cite{GoA92} where for the first time a
consistent treatment including leading order relativistic contributions
had been presented for a pure one-pion-exchange model.
As mentioned above, we have extended this work to the realistic Bonn OBEPQ
versions and first applied it to electrodisintegration \cite{FR97}
where further details can be found.
In particular, all 
explicit expressions for the electromagnetic operators, which
can be derived in the equation-of-motion method \cite{GoA92} or in the
unitarily equivalent $S$-matrix approach \cite{AT89},
are listed in the Appendix of \cite{FR97}, 
including in addition the 
$\gamma\pi\rho$- and $\gamma\pi\omega$-currents and the
currents involving $\Delta$-isobars.
These dissociation and isobar currents introduce additional uncertainties
and model dependence and will not be considered in the present study.

When calculating electromagnetic properties of hadronic systems one must,
of course, use a set of electromagnetic operators that is consistent with
the underlying hadronic interaction model, as is demanded by the
requirement of gauge invariance. 
In order to construct such a consistent set of
operators one should use the same theoretical basis as for the hadronic 
interaction, i.e., for
a one-boson-exchange potential (OBEP) one should calculate all 
corresponding meson exchange current (MEC) operators consistently. 
However, it is not sufficient to simply take the same meson
coupling constants and cutoffs as used in the potential model. 
Especially for the pion, care must be taken, because
there are several sources of unitary freedom in
constructing the pionic operators, leading to unitary parameters
in the corresponding current expressions \cite{GoA92,AT89}. 
These should be chosen consistently with the OBE potential and for this reason
we will briefly discuss them now.

First of all, in view of the unitary equivalence of pseudoscalar (ps)
and pseudovector (pv) pion nucleon couplings,
one introduces a mixing parameter $\mu$, that allows arbitrary
mixing of the two coupling types, 
where $\mu=0$ corresponds to pure ps
and $\mu=1$ to pure pv coupling.
Secondly, it is well known, that this equivalence breaks down
in the presence of the electromagnetic interaction.
Then one has the choice between a ps-pv equivalent theory or one which is not,
i.e.\ one can retain or leave out the equivalence breaking term.
Clearly, chiral symmetry prefers the pv coupling.
Thus it is customary to multiply the equivalence breaking term with a 
new parameter $\nu$ that switches this term off ($\nu=0$) or on ($\nu=1$)
and introduces as additional parameter 
\begin{equation}
\gamma=\mu+\nu,
\end{equation}
so that $\gamma=1$ corresponds
to a chiral invariant interaction theory whereas
$\gamma=0$ violates chiral symmetry.

Another freedom arises in the $p/M$ expansion for
the nonrelativistic reduction, 
the so-called Barnhill freedom described by
the Barnhill parameter $c$ \cite{Barnhill}, 
which can be incorporated in the 
definition of a parameter
\begin{equation}
  \tilde{\mu}=\mu+c(1-\mu).
\end{equation}
Note that only for ps coupling one has a dependence on the Barnhill parameter.

Finally, another unitary parameter stems from retardation. Although the OBEPQ 
versions are static potentials, it is not possible to construct a gauge 
invariant set of electromagnetic operators that are purely static \cite{AT89}. 
This is due to the non-local nature of the e.m.\ operators when one leaves
the nonrelativistic limit, e.g., the charge density associated with the
pion in flight fulfills the gauge condition with the other retarded operators
\cite{GoA92}. We have  generated the retarded potential through a Taylor
expansion of the pion propagator keeping the leading term only
\begin{equation} 
V_{ret}(\vec{k}) = V_{0}(\vec{k})\Delta(\vec{k}^2) k_0^2, 
\end{equation}
where $V_{0}(\vec{k})$ is the static, nonrelativistic potential, $k_0$
the energy transfer at the vertex, and
\begin{equation}
\Delta(\vec{k}^2) = \frac{1}{m^2+\vec{k}^2} 
\end{equation}
the static meson propagator. 
Certainly, this approximation is valid only
below the pion production threshold.
Here one has
again the freedom to express the energy transfer of the pion by the
energy transfers of the individual nucleons parametrized by a
retardation parameter $\nu_{ret}$
\begin{eqnarray}
 k_0^2 &=& -k_0^{(1)}k_0^{(2)}+
 \frac{1-\nu_{ret}}{2}(k_0^{(1)}+k_0^{(2)})^2 \nonumber\\
  &=& \frac{1}{4M^2}\left(\vec{k}\!\cdot\!\vec{Q}_1\vec{k}
\!\cdot\!\vec{Q}_2
   + \frac{1-\nu_{ret}}{2}\left(\vec{k}\!\cdot
   \!(\vec{Q}_1-\vec{Q}_2)\right)^2\right),
\end{eqnarray}
where $k^{(i)}_0$ denotes the energy transfer of nucleon ``$i$''.
This freedom can be
exploited to eliminate the retarded potential in the center-of-mass (c.m.)
frame by the choice $\nu_{ret}=\frac{1}{2}$.
The retarded e.m.\ operators
must then be constructed consistent with this choice.

With respect to the Bonn OBEPQs, 
one must be aware that these were constructed from the
three dimensional Blankenbecler-Sugar reduction of the
Bethe-Salpeter equation, thus yielding the nonrelativistic form of the
kinetic energy operator, while still being a fully relativistic potential.
The e.m.\ operators, constructed within the time-ordered
perturbation theory or within an equation-of-motion approach \cite{GoA92},
contain naturally relativistic one-body operators. The connection between
potential operators, calculated within these two different approaches,
 can be made by the ``Coester'' transformation \cite{AT89}.
This leads to the conclusion that the operators in \cite{AT89} are consistent
with the Bonn potentials for the choice $\tilde{\mu}=-1$, as was found
in \cite{GoA92}.

For the exchange of heavy mesons, the operators given in \cite{AT89} are
consistent with the Bonn potentials
due to the simplicity of the corresponding Hamiltonians in leading
order. 
For this reason, no additional unitary parameters appear. They should show
up in higher order terms, which however, can safely be neglected due to the
large meson masses.

For the $\rho$ MEC we had distinguished in \cite{FR97} between a
``Pauli'' current, 
corresponding to the $\rho$ contribution proportional to $(1+\kappa_V)^2$
which can be generated from the $\pi$ MEC by substituting terms of the form 
$\vec{\sigma}\cdot\vec{a}$ by $\vec{\sigma}\times\vec{a}$, 
and the ``Dirac'' current for the remaining operators. 
This distinction will also be used later in the discussion.

With respect to the $\sigma$ meson, we would like to mention
that the Bonn potentials OBEPQ (A,B,C) need different meson parameters for 
the isospin $T=0$ and $T=1$ channels. 
Introducing the isospin projection operators,
the potential (and the e.m.\ operators) can be viewed as a superposition 
of an isoscalar and isovector scalar
meson for the two sets of parameters, yielding effectively four scalar mesons
\begin{equation}
  V^{\sigma} =
\frac{1}{4}\left(1-\vec{\tau}_1\!\cdot\!\vec{\tau}_2\right)
V^{\sigma_0}
 +\frac{1}{4}\left(3+\vec{\tau}_1\!\cdot\!\vec{\tau}_2\right)
V^{\sigma_1}.
\end{equation}
So, strictly speaking, one has to take into account 
artificial isovector currents introduced through this procedure, which
obviously tend to cancel each other. A simpler approximation, i.e.,
comparing the ``pure'' $\sigma$ meson exchanges of the two parameter sets,
also leads to the conclusion that this ``anomaly'' has no visible
effect \cite{Levchuk}.

Finally, for a consistent treatment of leading order relativistic
contributions one has to include the wave function boost which
takes into account the fact that initial and final hadronic states move
with different velocities, i.e., their rest frames to which the
intrinsic motion refers are different.
A convenient method for treating this frame dependence of the 
intrinsic motion is to introduce a
unitary transformation generated by a boost operator $\chi(\vec{P})$
\begin{equation}
\mid{\vec{P},\vec{p}}\ \rangle =
\mid{\vec{P}}_{c.m.}\ \rangle \otimes
 \mbox{e}^{-i\chi(\vec{P})}\mid{\vec{p}}_{int}\ \rangle,
\end{equation}
where 
$\mid{\vec{p}}_{int}\ \rangle$ describes the intrinsic wave
function in the rest frame \cite{KrF74}.
Instead of transforming the wave functions themselves, the boost effect is
incorporated into the operators by
\begin{equation}
 \mbox{e}^{i\chi} \Omega
 \mbox{e}^{-i\chi}
 \approx
 \Omega+i\left[\chi,\Omega\right],
\end{equation}
where in the commutator only the nonrelativistic part of $\Omega$ has to be
considered.

The operator $\chi$ can be separated into a kinematic, interaction
independent part $\chi_0$ and an interaction dependent part $\chi_V$
\begin{equation}
  \chi=\chi_0 + \chi_V.
\end{equation}
For the two nucleon system one has \cite{KrF74}
\begin{equation}
\label{boost.krf}
\chi_0 = -\left( 
 \frac{(\vec{r}\!\cdot\!\vec{P})
   (\vec{p}\!\cdot\!\vec{P})}{16M^2} + h.c. \right)
 + \frac{((\vec{\sigma}_1-\vec{\sigma}_2)\!\times\!\vec{p})
 \!\cdot\!\vec{P}}{8M^2},
\end{equation}
whereas
a nonvanishing, interaction dependent boost operator exists only for
pseudoscalar meson exchange ($\pi$, $\eta$) in the case of
pseudoscalar coupling \cite{Fri77,Fri79}
\begin{equation}
 \chi^\pi_V = -(\vec{\tau}_1\!\cdot\!\vec{\tau}_2)
 \frac{i}{8M} 
 \left(\frac{{g_{\pi NN}}}{2M}\right)^2
 (1-\tilde{\mu})
 \int\!\!\frac{d^3k}{(2\pi)^3}
 \mbox{e}^{i\vec{k}\cdot\vec{r}}\Delta(\vec{k}^2)
 \vec{\sigma}_1\!\cdot\!\vec{P} \vec{\sigma}_2\!\cdot\!\vec{k}
 +( 1\!\leftrightarrow\! 2 ).
\end{equation}
Therefore, for the Bonn potentials, the potential dependent boost
appears with $(1-\tilde{\mu})=2$.

\section{Results and Discussion}
\label{results}

We have calculated the unpolarized differential cross section 
$\frac{d\sigma_0}{d\Omega}$ and all single polarization observables, i.e.,
photon asymmetry $\Sigma^l$, the target asymmetries $T_{11}$, 
$T_{20}$, $T_{21}$, $T_{22}$, and the final nucleon polarization $P_y$
for proton and neutron.
Their formal expressions in terms of the basic T-matrix elements
\begin{equation}
  T_{s m_s \lambda m_d} =
  \pi \sqrt{\alpha\frac{k}{q} E_d}
  \langle s m_s \mid j_\lambda(\vec{q})
  \mid m_d \rangle
\end{equation}
are given in \cite{HA88},
where $\alpha$ denotes the fine structure constant,
$k$ the asymptotic relative momentum of the outgoing nucleons in the 
c.m.\ frame,
$q$ the photon momentum,
$E_d$ the energy of the initial deuteron, 
and $j_\lambda$ the nucleon current in a spherical basis.

For the explicit calculation we have chosen four representative photon energies
$E_\gamma=4.5$, $40$, $100$, and $140$ MeV covering the region between the
maximum of the total cross section and pion production threshold.
In order to distinguish the different influences from pion, rho, and other
heavy meson exchanges, we show their effects in separate panels for each 
observable and each energy. In addition, we show an overview and the potential
model dependence with respect to the versions A, B, and C of the Bonn OBEPQ\@.
Thus each observable is represented by a figure consisting of four columns,
one for each energy, and five rows for the overview, $\pi$ exchange, $\rho$
exchange, additional heavy meson exchange, and potential model dependence.
The notation of the curves is the same for all observables.

In the first row we present an overview of the following effects:
the nonrelativistic one-body
current (long-dashed curve), the relativistic one-body current
(dash-dotted curve), to this added
the  nonrelativistic $\pi$ MEC (dotted curve), 
and finally the total result including all heavy meson exchanges
(full curve).

The next row shows the contributions from $\pi$ MEC, 
starting from the relativistic one-body current 
(dash-dotted curve of the first row) to which first
the nonrelativistic $\pi$ MEC is added (dotted curve)
and then the relativistic 
$\pi$ MEC -- including retardation corrections -- (dashed curve).

The third row displays the effects of the $\rho$ MEC: here we start from 
the relativistic $\pi$ MEC (dashed curve of the previous row) and include
first the Pauli $\rho$ MEC (short-dashed curve) and then the
Dirac $\rho$ MEC (long-dashed-dotted curve).

The effects of the various heavy meson exchanges are presented in the
fourth row. 
To the relativistic one-body current plus relativistic $\pi$ MEC and full 
$\rho$ MEC (long-dash-dotted curve, as in the third row), 
we add consecutively
$\delta$ MEC (dotted curve),
$\omega$ MEC (short-dashed curve),
$\sigma$ MEC (dashed curve),
and finally the $\eta$ MEC (full curve).

The last row shows the potential dependence of the observable
with respect to the different versions of the Bonn OBEPQ potential,
where the
full curve represents the version B, 
the short-dashed version A, 
and the dotted version C.

\begin{figure}
\centerline{%
\epsfxsize=16.50cm
\epsffile{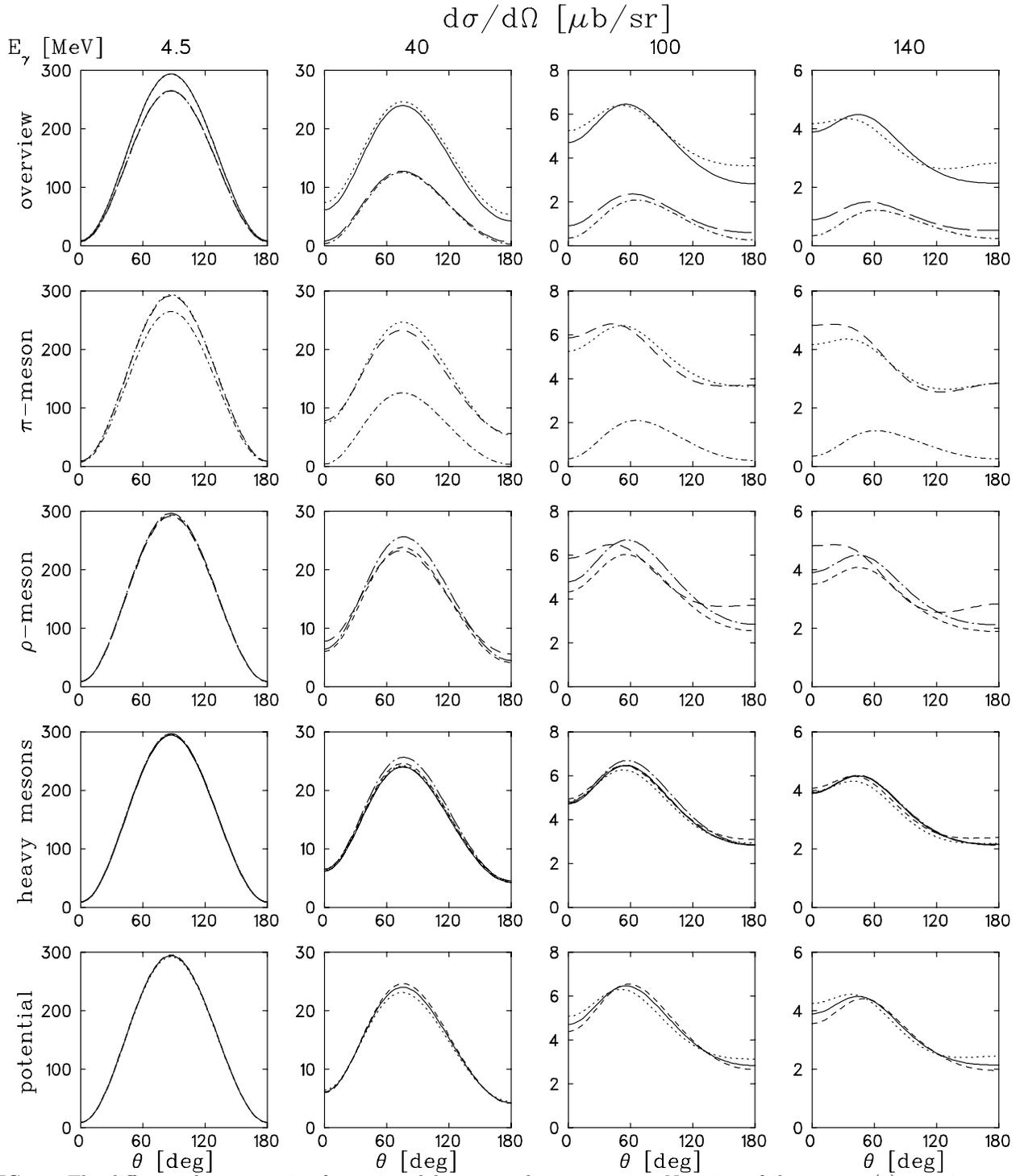}
}
\caption{%
\label{diffWq}
The differential cross section for various laboratory photon energies.
Notation of the curves:
(1) overview: 
nonrelativistic one-body current (long-dashed);
relativistic one-body current (dash-dotted);
nonrelativistic $\pi$ MEC added (dotted); 
total result (full);
(2) $\pi$ meson:
relativistic one-body current (dash-dotted);
nonrelativistic $\pi$ MEC added (dotted);
relativistic $\pi$ MEC including retardation (dashed);
(3) $\rho$ meson:
relativistic one-body plus complete $\pi$ MEC (dashed);
Pauli MEC (short-dashed);
Dirac MEC (long-dashed-dotted);
(4) heavy meson:
relativistic one-body current plus complete $\pi$ and full 
$\rho$ MEC (long-dash-dotted);
$\delta$ MEC (dotted);
$\omega$ MEC (short-dashed);
$\sigma$ MEC (dashed);
$\eta$ MEC (full);
(5) potential:
OBEPQ version B (full);
version A (short-dashed);
version C (dotted).
}
\end{figure}

Now we will discuss in detail the various observables starting with the 
differential cross section in Fig.~\ref{diffWq}. 
The overview shows that in the maximum of the total cross section,
at $4.5$ MeV 
this observable
is dominated by the nonrelativistic
one-body current while only the nonrelativistic $\pi$ MEC gives a 10 percent 
enhancement. Obviously relativistic effects and heavy mesons are negligible 
as well as the potential model dependence. 

At the next higher energy ($40$ MeV) the nonrelativistic $\pi$ MEC becomes 
comparable to the one-body current. All other contributions give a small 
overall reduction, somewhat more pronounced in forward and backward 
direction. However, if one looks at the separate contributions, one notices 
a subtle destructive interference of different larger effects. First, 
relativistic $\pi$ MEC gives a slight reduction in the maximum but leaves 
the forward and backward directions almost unchanged. Next, from $\rho$ one 
sees a strong forward and backward reduction from the Pauli current, 
whereas the Dirac contribution mainly leads to a sizeable enhancement in 
the maximum which, however, is largely cancelled by the additional heavy 
mesons. Finally, one finds a small model dependence of a few percent. 

Considering now the two higher energies ($100$ and $140$ MeV) one readily 
notices a dramatic increase of relativistic effects. First a sizeable 
reduction appears from the relativistic one-body current showing the well known 
effect of diminishing the differential cross section at forward and 
backward angles which comes mainly from the dominant spin-orbit (SO) current 
\cite{Cam82}. The further reduction from the remaining contributions shows 
again in detail a strong destructive interference. In fact, first the 
relativistic $\pi$ MEC surprisingly enhances the cross section in forward 
direction while then the $\rho$ Pauli current results in a drastic reduction 
at both extreme angles. The Dirac contribution gives again an overall 
enhancement but of smaller size. 
This effect of the Dirac $\rho$ current is somehow surprising, because it is
roughly of the same size as that of the Pauli current whereas from the size 
of the coupling constants one would have expected a suppression by a factor 
of about $50$. The additional heavy mesons beyond the 
$\rho$ show a much smaller influence. 
Their individual contributions are surprisingly big (up to $\sim{}5$\%), 
in particular compared to their role in the parametrization of the $NN$ force 
and their importance in the electrodisintegration of the deuteron.
Most prominent is the effect of the $\delta$ meson leading to a reduction of
the differential cross section. However, looking at the overall result, 
the heavy meson exchanges tend to cancel each other. 
With respect to the potential model 
dependence, one sees now a larger variation, in particular also at forward 
and backward angles which increases with energy.

\begin{figure}
\centerline{%
\epsfxsize=16.50cm
\epsffile{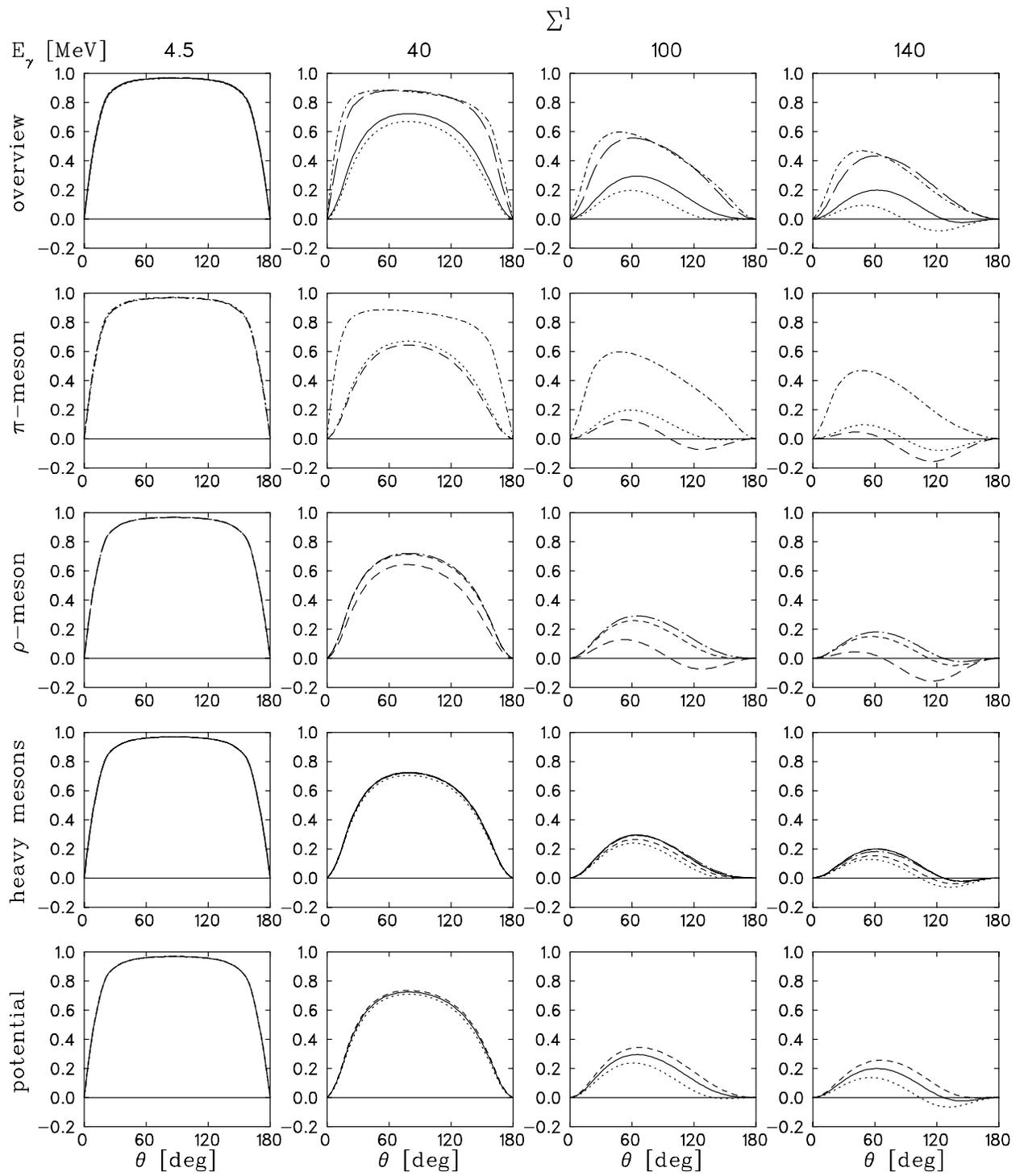}
}
\caption{%
\label{sigma}
The photon asymmetry $\Sigma^l$.
Notation of the curves as in Fig.~\protect{\ref{diffWq}}.
}
\end{figure}

The photon asymmetry $\Sigma^l$ in Fig.~\ref{sigma} is very
sensitive to two-body effects, as is long known \cite{ArS91}. 
At $4.5$ MeV only the nonrelativistic one-body current contributes to 
$\Sigma^l$ and no potential model dependence appears. Then at $40$ MeV 
the nonrelativistic $\pi$ MEC becomes sizeable as well as the Pauli $\rho$ 
current. All other effects, relativistic one-body and $\pi$ MEC, Dirac 
$\rho$ and additional heavy meson effects are very small, as is the 
potential model dependence. 
At higher energies the relativistic one-body current 
and the relativistic $\pi$ MEC becomes important, too.
The first leads to a sizeable reduction of the photon asymmetry, 
the latter to a smaller increase. The $\rho$ MEC increases the photon 
asymmetry, of which the Pauli current is the most 
dominant part while the Dirac current is comparably small, although its size
increases with the photon energy. 
The influence of the various heavy meson
exchanges are much more pronounced than in the differential cross section,
mainly coming from the $\delta$ MEC\@. But again the various heavy mesons tend
to interfere destructively. 
Also the potential dependence is quite large for $100$ MeV and $140$ MeV,
where the OBEPQ version A yields the biggest asymmetry, version B intermediate
values, and version C the lowest photon asymmetry. Thus one might be tempted
to single out one potential against others by comparison with experimental
data. However, one has to be careful in such a comparison \cite{blanpied},
because before drawing definite conclusions as to which model should be
preferred, one has to study in detail the remaining theoretical uncertainties
due to the strength of the dissociation and isobar currents. Here, the
additional independent measurement of the unpolarized cross section and the
vector target asymmetry $T_{11}$ could help in fixing the respective strengths
of these contributions.

\begin{figure}
\centerline{%
\epsfxsize=16.50cm
\epsffile{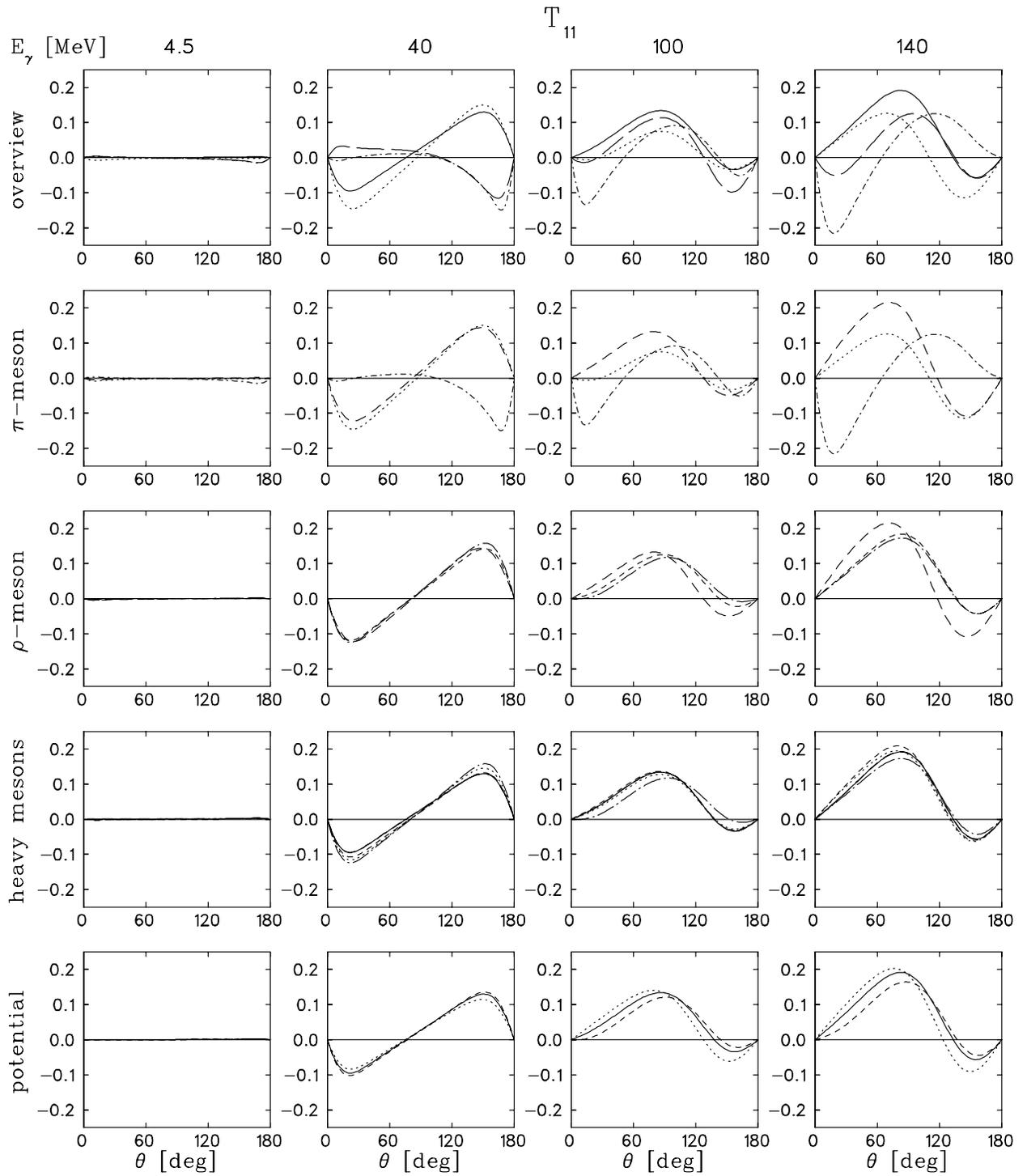}
}
\caption{%
\label{t11}
The vector target asymmetry $T_{11}$.
Notation of the curves as in Fig.~\protect{\ref{diffWq}}.
}
\end{figure}

The vector target asymmetry $T_{11}$, shown in Fig.~\ref{t11}, 
is extremely small at $4.5$ MeV and thus will not be discussed furthermore. 
However, at $40$ MeV, $T_{11}$ develops a strong sensitivity
to the relativistic one-body current as well as to the nonrelativistic and
relativistic $\pi$ MEC leading to an overall sign change and a strong 
increase in absolute size around 30 degree. The $\rho$ MEC shows only 
small effects, slightly more pronounced for the heavier mesons. Also the 
variation with the potential model remains marginal. 
For the two higher photon energies one readily sees an even more dramatic 
influence of the various currents. Already the relativistic one-body 
current shows a strong effect, which is partly cancelled from 
nonrelativistic and relativistic $\pi$ MEC\@. Also visible effects come from 
the $\rho$ MEC, mainly Pauli, and from the heavy meson sector, again mainly 
from the $\delta$. 
The potential dependence is as big as for the photon asymmetry $\Sigma^l$.

\begin{figure}
\centerline{%
\epsfxsize=16.50cm
\epsffile{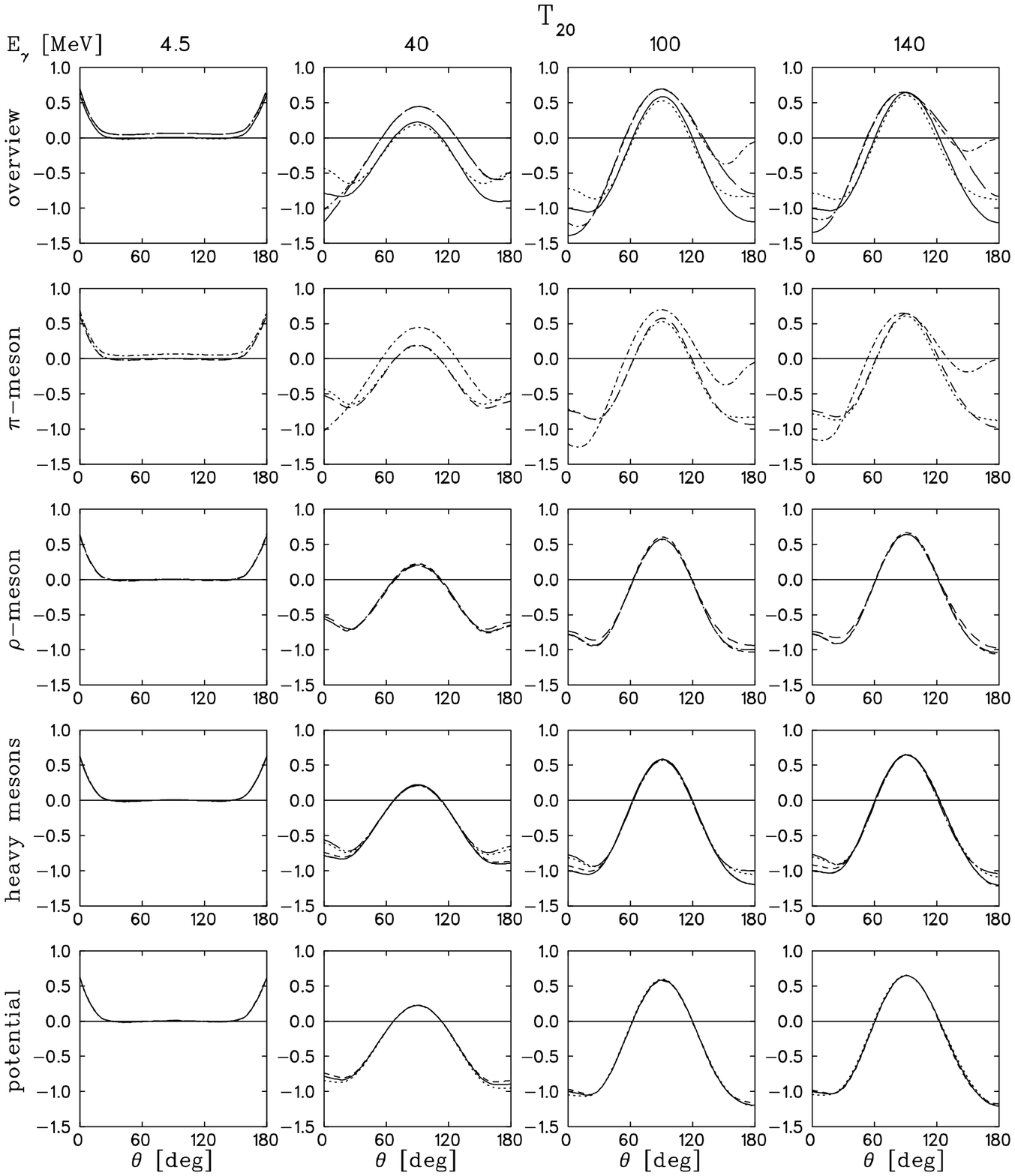}
}
\caption{%
\label{t20}
The tensor target asymmetry $T_{20}$.
Notation of the curves as in Fig.~\protect{\ref{diffWq}}.
}
\end{figure}

The next observable, the tensor target asymmetry $T_{20}$ in 
Fig.~\ref{t20} shows sizeable effects from the various currents mainly for the 
regions around $0^\circ$ and $180^\circ$, except at $4.5$ MeV where only a 
very small influence from the nonrelativistic $\pi$ MEC is seen. For the 
higher energies, the largest effect stems from the nonrelativistic $\pi$ 
MEC\@. In addition, at $100$ and $140$ MeV also 
the relativistic one-body current 
becomes equally important, particularly strong in the backward direction of 
increasing size.  
This observable reacts only slightly when the relativistic $\pi$ MEC is 
added, and the same is true for the $\rho$ current. The additional heavy 
mesons appear more interesting. The $\omega$ and $\sigma$ meson exchanges 
lead to a visible reduction at forward and 
backward angles. Finally, $T_{20}$ is stable against a change of 
the potential version. This is valid for all tensor target asymmetries.

\begin{figure}
\centerline{%
\epsfxsize=16.50cm
\epsffile{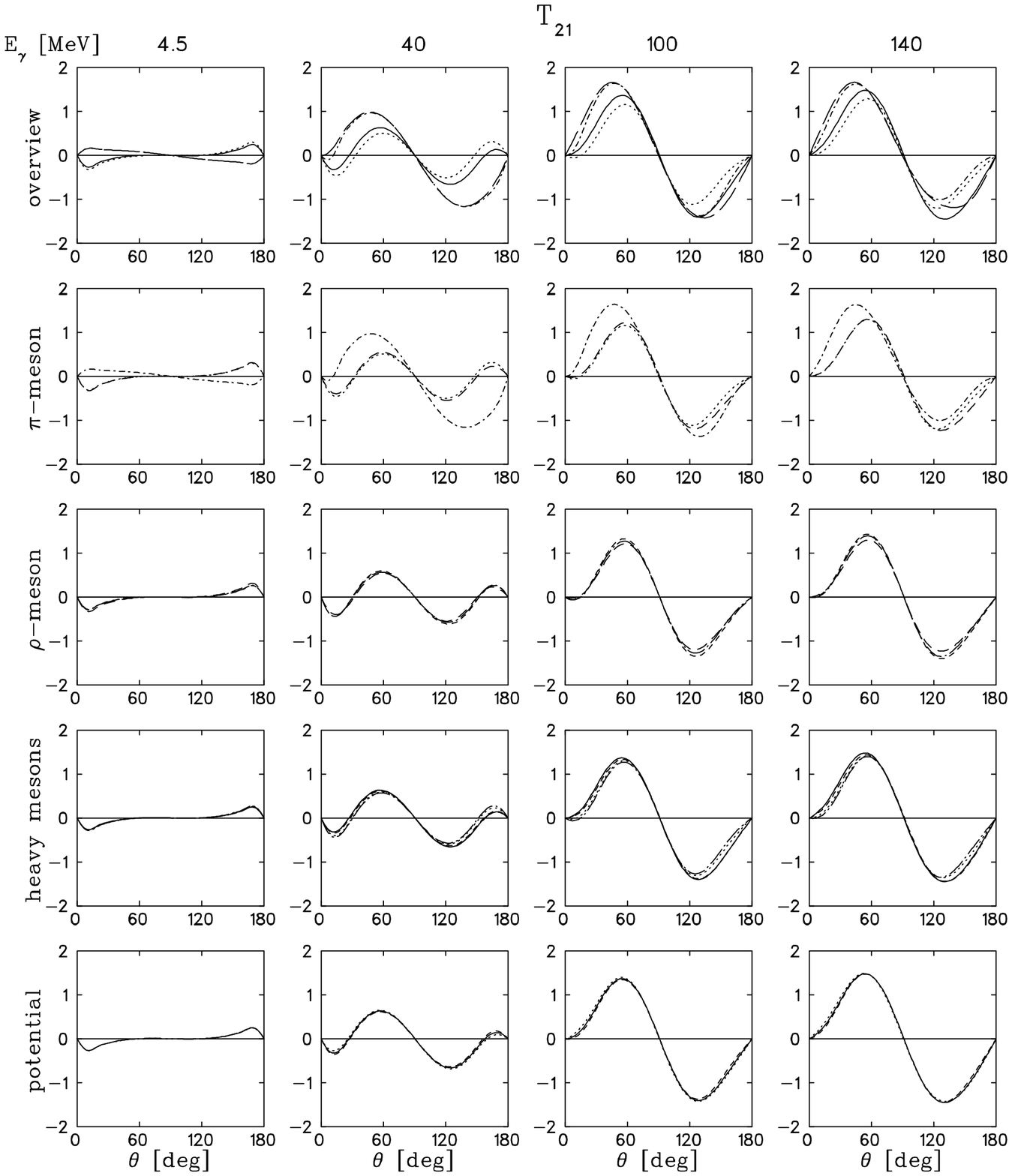}
}
\caption{%
\label{t21}
The tensor target asymmetry $T_{21}$.
Notation of the curves as in Fig.~\protect{\ref{diffWq}}.
}
\end{figure}

The next tensor target asymmetry $T_{21}$ in Fig.~\ref{t21} shows the 
largest sensitivity to the nonrelativistic $\pi$ meson exchange at the low  
energy of $4.5$ MeV, although the observable is small in absolute size. Also 
for the higher energies the only relevant contribution besides the one-body 
part comes from the nonrelativistic $\pi$ MEC, but the relative importance 
diminishes with increasing energy. Otherwise, $T_{21}$ is a very stable 
observable with respect to all remaining contributions and to a potential 
variation.

\begin{figure}
\centerline{%
\epsfxsize=16.50cm
\epsffile{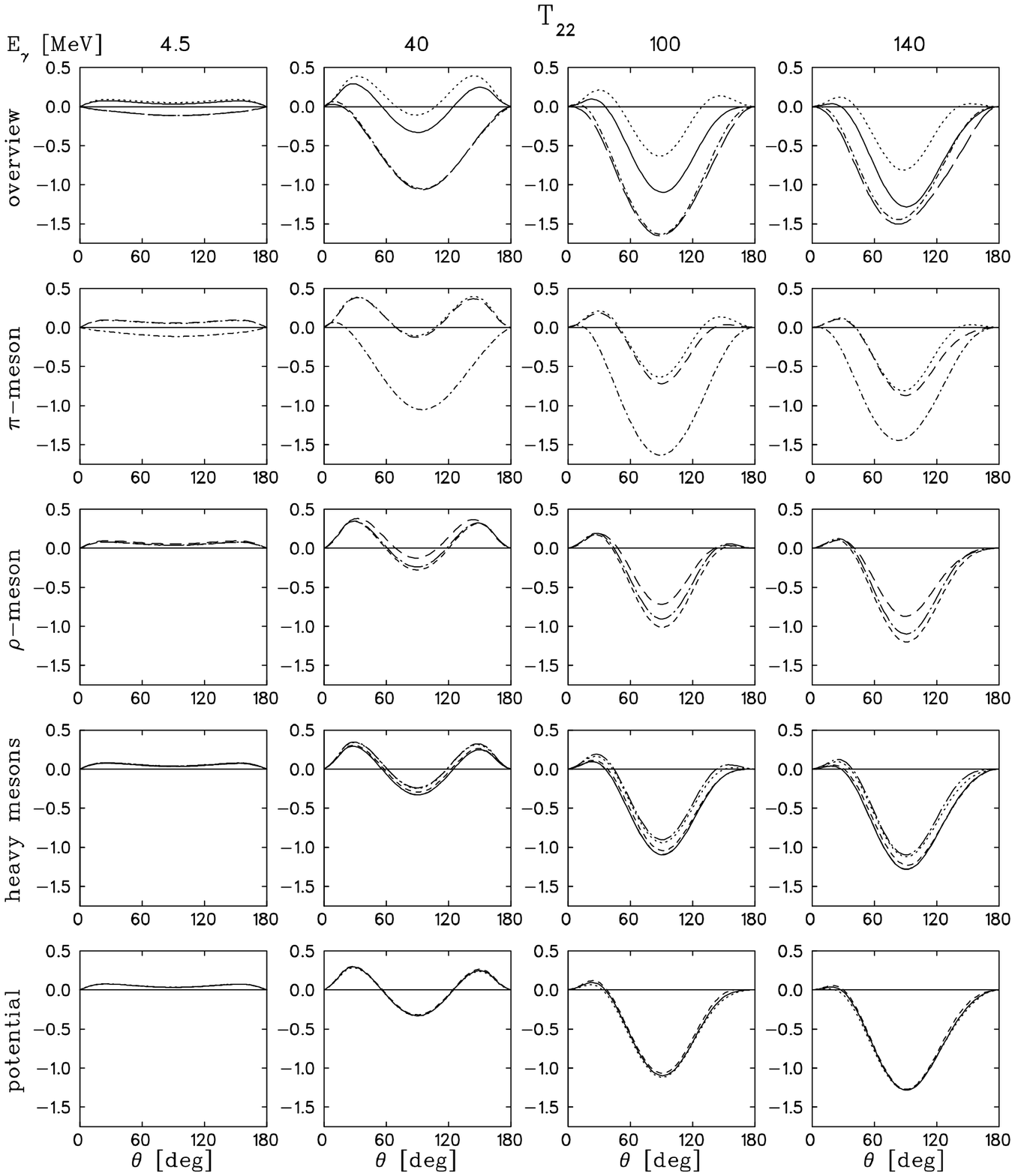}
}
\caption{%
\label{t22}
The tensor target asymmetry $T_{22}$.
Notation of the curves as in Fig.~\protect{\ref{diffWq}}.
}
\end{figure}

The last tensor target asymmetry $T_{22}$ in Fig.~\ref{t22}
is like $T_{21}$
also sensitive to two-body effects of which the 
nonrelativistic $\pi$ MEC is dominating again.
It leads for photon energies of $40$, $100$, and $140$ MeV to a drastic
increase of the asymmetry. But this observable shows almost no influence 
from relativistic contributions, neither from the relativistic one-body nor 
from the relativistic $\pi$ MEC\@. Also the $\rho$ current shows an 
increasing contribution with increasing energy, mainly from the Pauli part. 
The additional heavy meson exchange becomes significant, too, with the 
largest part from $\omega$ MEC, leading to a slight overall decrease. The 
variation with the potential model is negligible.

\begin{figure}
\centerline{%
\epsfxsize=16.50cm
\epsffile{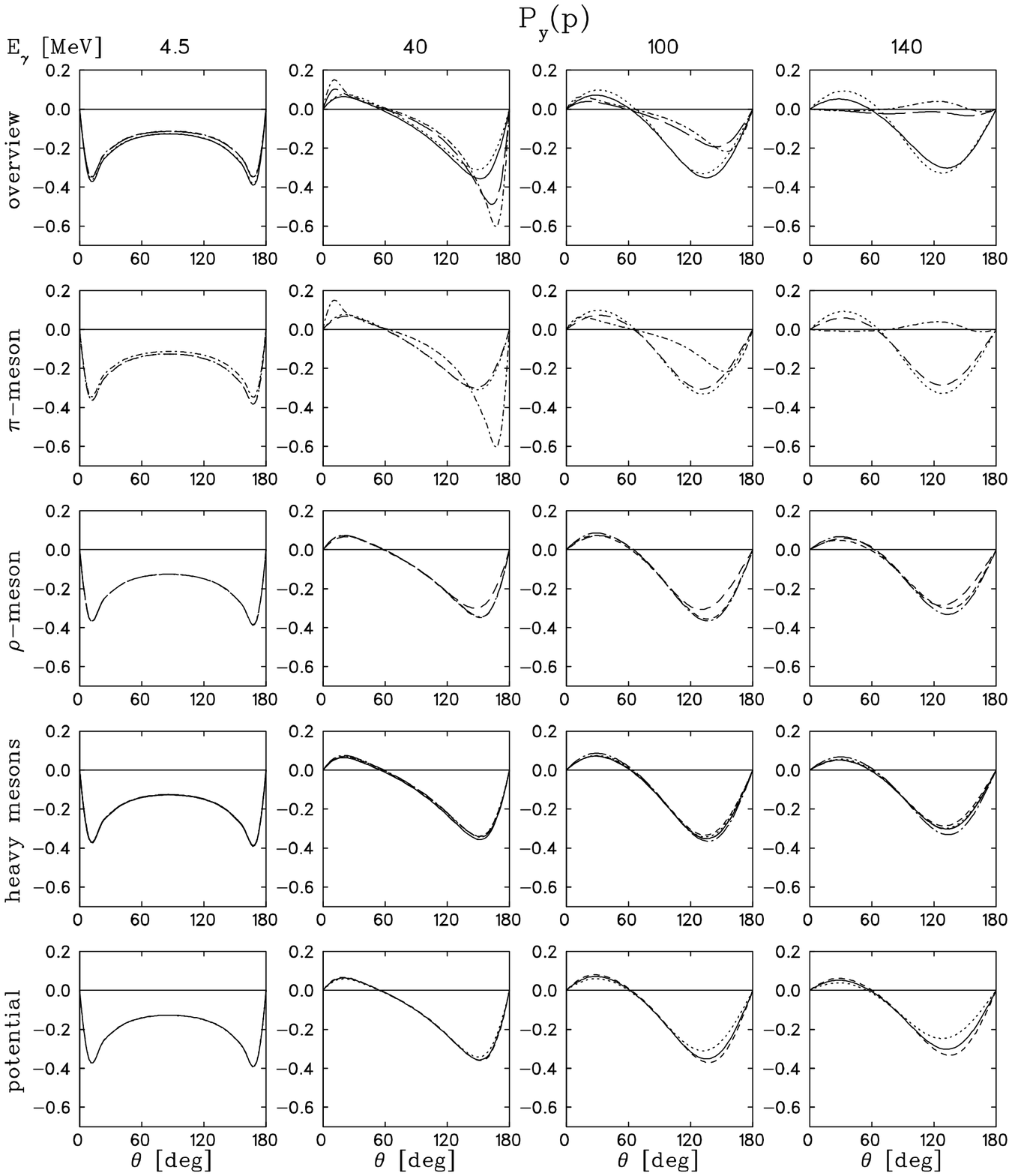}
}
\caption{%
\label{pyp}
The polarization $P_y$ of the outgoing proton.
Notation of the curves as in Fig.~\protect{\ref{diffWq}}.
}
\end{figure}

\begin{figure}
\centerline{%
\epsfxsize=16.50cm
\epsffile{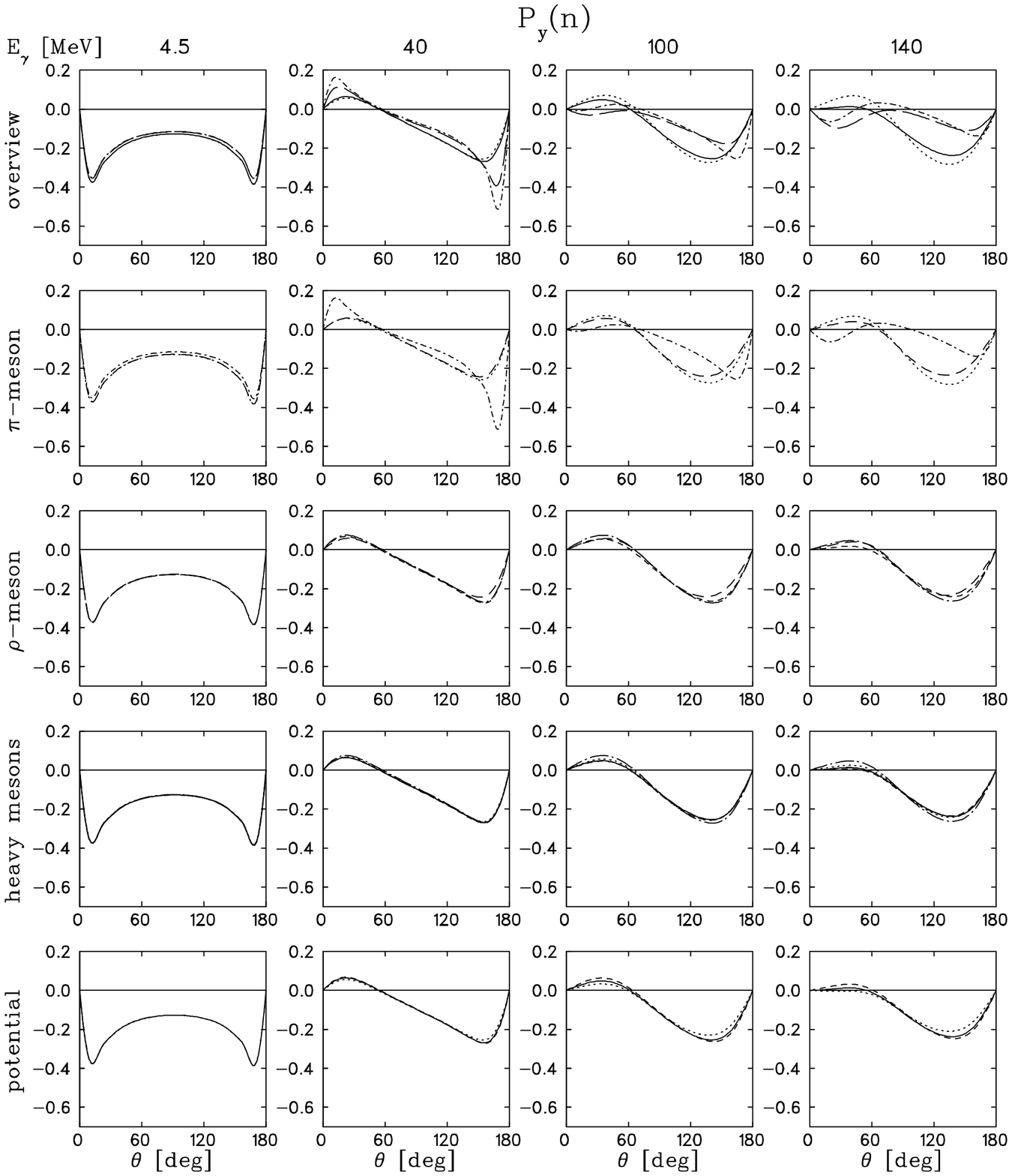}
}
\caption{%
\label{pyn}
The polarization $P_y$ of the outgoing neutron.
Notation of the curves as in Fig.~\protect{\ref{diffWq}}.
}
\end{figure}

Finally, we show the outgoing proton and neutron polarizations $P_y(p)$ and 
$P_y(n)$ in Figs.~\ref{pyp} and \ref{pyn}, respectively. They show very 
similar behaviour with respect to the different contributions, except 
for the two higher energies where one notes some differences.
At $4.5$ MeV the polarization is dominated by the nonrelativistic one-body 
current. There is only a tiny contribution from the nonrelativistic $\pi$ 
MEC, and all other effects are completely negligible. The relativistic 
one-body current shows some increasing influence with increasing energy, 
but the most important contribution is here again the nonrelativistic 
$\pi$ MEC whereas the sensitivity to the relativistic $\pi$ MEC is very 
small. The $\rho$ contribution stems mainly from the Pauli $\rho$ MEC,
leading to a reduction at backward angles. The additional heavy meson 
exchanges lead to a small but increasing effect for higher photon energies. 
The potential dependence becomes sizeable at $100$ and $140$ MeV, 
in particular for the proton polarization, and also increases with energy.

Finally, we would like to mention that boost contributions have turned out 
to be totally negligible. This is true for both the one-body boost operators, 
see also \cite{GoA92}, as well as for the contributions of the 
two-body boost currents. It is in contrast to deuteron electrodisintegration 
where the one-body boosts cannot be neglected \cite{WiB93} whereas the 
two-body boost is small also there \cite{FR97}.

\section{Summary and Outlook}
\label{summary}

Summarizing our results, we may state that relativistic contributions 
beyond the leading spin orbit current are important and should be 
included in a consistent manner within a realistic hadronic interaction model. 
In comparison with the work of Ref.~\cite{GoA92},
where a pure pionic model had been used, 
we see that the relativistic effects are roughly of the same size.
For the differential cross section one notices
that the relativistic pionic current
in the present work has a different angular distribution than in \cite{GoA92},
whereas the effects in the photon asymmetry are very similar.
These differences stem from the different potential models 
and were expected.

Of the heavy mesons beyond the pion, the $\rho$ meson is the most important 
one confirming earlier investigations. We also find that in many cases the 
Pauli current is dominating its contribution justifying to some extent 
the neglect of the Dirac current. However, we have found some observables 
for which the Dirac current gives contributions comparable to the Pauli 
current. 
With respect to the other heavy mesons beyond the $\rho$ meson, 
we have found that each individual meson shows a sizeable influence 
on the various observables of deuteron photodisintegration.
But taken together, we find in many of the observables considered strong 
destructive interference between them 
($\delta$, $\omega$, $\sigma$, and $\eta$)
so that then their overall effect is quite small.
Nonetheless, there are certain polarization observables, 
like the tensor target asymmetry $T_{20}$,
that obviously demand the incorporation of the 
heavy meson contributions.

We have not attempted to make a comparison with experimental data because, 
first of all, our main interest was the study of the relative importance of 
relativistic effects for a realistic $NN$ interaction model compared 
to the current 
contributions from heavy meson exchange. Secondly, since at low and medium 
energies the results for cross sections and polarization observables agree 
essentially with those of other potential models, one can expect the same 
kind of agreement with experimental data for these energies, whereas at 
higher energies ($100$ and $140$ MeV) isobar contributions, 
particularly from the $\Delta$ resonance, 
have to be included for a meaningful comparison. 
Furthermore, above pion production threshold, the
approximative treatment of retardation in the present work breaks down 
and the correct incorporation of pion retardation in the $NN$- and
$N\Delta$-sector becomes important \cite{MSchwamb}. These effects
are definitively much more crucial than the effects of heavy meson 
exchange. Also relativistic contributions to the excitation of the 
$\Delta$ resonance itself might show some sizeable influence. 
These questions will be investigated in future work.

\end{document}